# LIVEWN: CPU SCAVENGING IN THE GRID ERA
*Expanding LCG/EGEE Grid infrastructure into the millions*

Koyretis G. <ge01275@mail.ntua.gr>, Georgatos F. <fotis@mail.ntua.gr>
*National Technical University of Athens, Greece*

Abstract:   This project's goal is to introduce an easy and versatile way to provide and use Grid resources without the need of any OS installation or middleware configuration. At the same time it provides an excellent training tool for newer Grid users and people that want to experiment, without enforcing installation. It has been tested thoroughly under different circumstances with firm success.

## INTRODUCTION

This project's primary goal is to add computational resources to the Grid [1] in a simple way, user friendly enough, that non-experts can use. So far, no one has accomplished fully dynamic resource management on the LCG/EGEE grid [2] [3], since there had to be coordination at the system administration level in order to tune installation parameters. The LiveWN image is a technology that solves this problem in a versatile way so that Worker Nodes (WNs) can be setup and used behind firewalls, within virtual machines, over dial-up lines, etc. It just works, without requiring expertise.

## RELATED WORK

There exist also projects that share similarities, like E-Grid and Gilda. The E-Grid livecd for example [4], while it works for some cases it has no provision for dynamic allocation of WNs and requires fully qualified hostnames and domainname (FQDNs) in order to work. On the other hand the GILDA project [5] has produced a live user interface (UI) but it has no provision of adding resources through the LiveCD (eg. supplying CPUs).

## TECHNOLOGY

Our solution is a mixture of three technologies, supplied by others:
- A Knoppix-like LiveCD [6]
- LCG/EGEE middleware [7]
- An OpenVPN IP tunnel [8]

Upon boot, the LiveWN is configured to ask an IP address with DHCP, and then it configures some initial network access parameters. Then, we have a script, which the user is asked to execute. The user is asked with login/password in order to "attach" to our Computing Element (CE); it is entirely possible to subscribe "anonymous" resources as well, but we don't want this feature enabled, yet. Once correct credentials have been supplied, an OpenVPN tunnel is created, and the system configures its hostname and domainname to be the proper ones, and boots in accordance to the associated Computing Element. Since we have indeed pre-configured correct forward and reverse DNS, it appears as just another Worker Node, so it joins the CE's queues and it starts accepting and executing jobs.

## TESTING

Our team tried the proposed solution in all following circumstances:
- A bare i386 PC system booting with a LiveWN disk; with success
- A 64bit Linux system running a Virtual Machine with qemu (VM/32bit), and then booting either by CD or local image; with success
- Windows XP system, running vmware and inside it a local image, again with success and an efficiency >90%; as measured for CPU-bound tasks.

We expect the performance of the LiveWN solution to be found up to par with what the Virtual Machines themselves can offer, so it should be good choice for a range of cases.

LiveWN is an adaptable technology, because the VPN technique allows it to run behind firewalls, on systems within private address space networks and, above all, under many unexpected platforms and environments.

## IMPLEMENTATION

The ability to tap computational resource from, say, Windows machines and adding them to the Grid, is something which allows the current LCG/EGEE testbed to jump easily into the numbers of hundreds of thousands of CPUs simply by employing cycle-scavenging techniques. There are plenty of applications that can benefit from this extra power, although surely not every application is fit (eg. weather models aren't, but folding proteins or analyzing radio-telescope signals are).

Our solution is to provide a LiveWN CD, which builds upon boot a VPN tunnel and attaches to a cluster's CE, which we supply upon initialization. All a user is asked to do is to run a script, type login/password, and even that step can be eliminated if required. It is extremely simple for an end-user, and very network-agnostic; it will run anywhere a 32 bit Linux operating system can boot and get an IP address; assuming surely it has resources to offer.

The impressive part of this work is that the amount of computing power that can become readily available for the LCG/EGEE testbed, which is currently the biggest one, are now multiplied to a level that perhaps is even an order or more of magnitude greater than currently, since equipment now is dedicated; assuming that most Windows users wouldn't like changing environment, too.

## FUTURE IMPROVEMENTS

What follows from now on is experimentation with the middleware software, gLite v3.0.X. We know it is not going to be easy, but neither has been LCG software, in the first place. It is though a necessary step in order to be ready for the next testbed. We are now under preparations to put this technology into a production service, which for us implies to reorganize our internal computing infrastructure, make more servers available so that we have redundant hardware for fail-over and load-balancing reasons etc. As a rule of thumb this technology performs optimally with applications that are able to do check-pointing; that is to save the intermediate state of their execution in "state" files, so that they can be restarted at a later point. This is, because resources on a scavenging grid can come and go very fast, so it is necessary to do some extra "housekeeping" in order to ensure job continuity. We have identified users within our University Department with applications that fall in this category and we are planning to assist them in using the LiveWN CD. For our own users' benefit we have added Povray software [9] on the disk, so that we can routinely perform distributed 3D rendering projects in the future.